\documentclass[journal,twocolumn]{IEEEtran}

\usepackage{color}
\usepackage{graphicx}
\usepackage{amsfonts}
\usepackage{amsmath}
\usepackage{graphicx}
\usepackage{cite}
\usepackage{epsfig}
\usepackage{latexsym}
\usepackage{subfigure}
\usepackage{epstopdf}
\usepackage{color}
\usepackage{verbatim}
\usepackage{units}
\usepackage{amsthm}
\usepackage{hyperref}
\usepackage[longend,ruled]{algorithm2e}

\begin{document}
\title{Ambient IoT: A missing link in 3GPP IoT Devices Landscape}

\author{M.~Majid~Butt,~Nitin R. Mangalvedhe,~Nuno K. Pratas, Johannes Harrebek, John Kimionis, Muhammad Tayyab, Oana-Elena Barbu, Rapeepat Ratasuk,~and~Benny Vejlgaard

\thanks{M. Majid Butt, Nitin R. Mangalvedhe and Rapeepat Ratasuk are with Nokia Standards, Naperville, USA. email:\{majid.butt, nitin.mangalvedhe, rapeepat.ratasuk\}@nokia.com.}
\thanks{Nuno K. Pratas, Johannes Harrebek, Oana-Elena Barbu, and Benny Vejlgaard are with Nokia Standards, Aalborg, Denmark. email: \{nuno.kiilerich\_pratas, johannes.harrebek, oana-elena.barbu, benny.vejlgaard\}@nokia.com.
}
\thanks{John Kimionis is with Nokia Bell Labs, 600 Mountain Ave, Murray Hill, NJ, 07974, USA (e-mail: ioannis.kimionis@nokia-bell-labs.com.)}

\thanks{Muhammad Tayyab is with Nokia Standards, Oulu, Finland. (e-mail: muhammad.tayyab@nokia.com.}
}

\maketitle
\begin{abstract}
Ambient internet of things (IoT) is the network of devices which harvest energy from ambient sources for powering their communication. After decades of research on operation of these devices, Third Generation Partnership Project (3GPP) has started discussing energy harvesting technology in cellular networks to support massive deployment of IoT devices at low operational cost. This article provides a timely update on 3GPP studies on ambient energy harvesting devices including device types, use cases, key requirements, and related design challenges. Supported by link budget analysis for backscattering energy harvesting devices, which are a key component of this study, we provide insight on system design and show how this technology will require a new system design approach as compared to New Radio (NR) system design in 5G.

\end{abstract}

\section{Introduction}
The internet of things (IoT) market is a major source of revenue for future wireless technology. Moving beyond enhancement of the mobile broadband devices market that is supported by fourth-generation (4G) cellular technology, ultra reliable low-latency communication (URLLC) and machine type communication (MTC) service classes were also introduced in fifth-generation (5G) cellular technology, which target use cases requiring high reliability and massive deployment of smaller and cheaper IoT devices, respectively. Sixth-generation (6G) technology will further expand these market segments. The IoT market is expected to steadily grow from 2020 to 2025, and further increase is expected beyond that period at a higher rate. Building and commercial, health, agriculture, infrastructure, and industries are some of the key sectors targeted by the IoT market. Almost all sectors are forecasted to grow in the future, but infrastructure and health segments are expected to grow at higher rates in the future compared to others \cite{report_AIoT}.

Due to the large number of IoT devices that are expected to be deployed, growth of the IoT market requires future networks to be energy efficient and sustainable to make a business case. Energy harvesting technology is key to the success of the IoT market as it can considerably reduce device operational cost. Energy harvesting devices are 'battery free' devices, which harvest energy from natural or ambient sources, e.g., electromagnetic, solar, thermal, pressure etc., and operate either with a small battery or without battery and do not require battery replenishment. They are also termed as 'zero energy' devices due to their capability to operate without a dedicated power source \cite{Lina}.

Energy harvesting communication has been widely discussed in research for more than a decade, e.g., see \cite{energy_harvestingsurvey, Wei_SWIPT} and references therein. Recent literature outlines key challenges and complexity to power massive number of IoT devices \cite{massive_WET,WET_ZE} using radio frequency (RF) signals, but this is the first time 3GPP has started discussing energy harvesting and backscattering as a potential IoT technology for future networks in various study groups \cite{AIoT_RANplenary, AIoT_SA}. Fundamental aspects of the technology are well researched, but feasibility of the technology in devices under 3GPP networks is still unknown under realistic network operating conditions and business model.

The goal of this article is to articulate the main reasons behind industry interest in energy harvesting as a future technology and discuss design considerations and solutions. As 3GPP has never before discussed energy harvesting devices, also called ambient IoT (A-IoT) devices, there are several challenges to be addressed before energy harvesting technology is adopted in future networks. This article provides a comprehensive overview addressing three key questions:
\begin{itemize}
  \item {How will energy harvesting IoT technology complement existing 3GPP IoT technologies?}
  \item {What are the key use cases and the associated design requirements for the technology?}
  \item {What are the challenges associated both with technology as well as its standardization in 3GPP?}
\end{itemize}

Section \ref{sect:IoT_space} discusses the first question by describing existing 3GPP IoT technologies and how energy harvesting technology fills the technology gap. Section \ref{sect:use_cases} addresses the second question by reviewing recent status of 3GPP discussions on energy harvesting technology and the way forward, while Section \ref{sec:sim} and Section \ref{sect:future} attempt to answer the third question by focusing on link-level analysis and the main technology challenges. Finally, we conclude with the main findings of the article in Section \ref{sect:conclusions}.
\begin{figure*}
\centering
  	\includegraphics[width=5.5in]{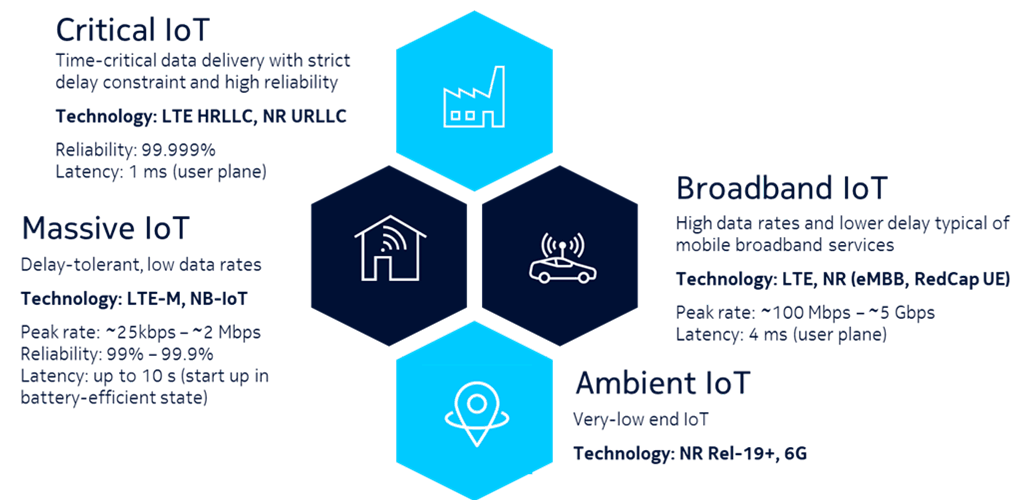}
   \caption{Cellular IoT Technology landscape with major technologies covering different potential use cases.}
	\label{fig:IoT}
\end{figure*}

\section{IoT SPACE IN 3GPP}
\label{sect:IoT_space}
The expansion of IoT to encompass use cases in new verticals has steadily progressed from the initial applications in second-generation (2G) cellular technology to the current 5G technologies. IoT use cases span a range of requirements in terms of cost, complexity, delay constraints, and other metrics.
Fig.~\ref{fig:IoT} depicts the division of IoT based on requirements into the following classes.\\
\textbf{Broadband IoT}: This class of IoT comprises one extreme of IoT applications that require high data rates along with a low latency. The use cases covered by this class of IoT include fleet management, industry gateways, video, hot-spots, wearables, and industrial wireless sensors. The requirements can be served by 4G Long Term Evolution (LTE) as well as 5G New Radio (NR) technologies (using enhanced Mobile Broadband, or eMBB, and Reduced Capability or RedCap, devices). RedCap devices have a lower cost and complexity than NR eMBB devices, making them more attractive for many IoT applications.\\
\textbf{Critical IoT:} This class of IoT constitutes another extreme of IoT applications requiring ultra-high reliability and ultra-low latency. Typical use cases for this class include factory automation, industrial control, robotics, and Augmented Reality (AR) or Virtual Reality (VR). The LTE High-Reliability Low-Latency Communication (HRLLC) feature of LTE, introduced in Release 15, and the NR URLLC technologies feature of NR, also introduced in Release 15, can serve these applications.\\
\textbf{Massive IoT:} This class of IoT consists of delay-tolerant applications where the requirements include low-cost devices with low energy consumption, extended network coverage, and support of a massive number of devices. Example use cases include fleet management, asset tracking, smart meters, smart city, gateways, sensors, voice, and point-of-sales. LTE for Machines (LTE-M) and Narrowband IoT (NB-IoT) technologies fulfill the requirements for these applications \cite{Rapeepat_NBIOT}.\\
\textbf{Ambient IoT:} The last class is the latest addition and comprises the very low end of IoT use cases where the requirements include ultra-low complexity devices, ultra-low power consumption, and small form factor, which can be met by devices that are either battery-less or have the capability for limited energy storage. The requirements related to power consumption and device cost/complexity cannot be met by existing cellular technologies. Therefore, new technologies under NR in Rel-19 or later and 6G are needed.

Fig.~\ref{fig:IoT} also shows some key performance indicators (KPIs) for the different cellular IoT technologies \cite{Nokia_RedCap}. It can be seen that in terms of peak data rate, broadband IoT is at the high end whereas massive IoT is at the low end. Among the legacy classes, in terms of reliability and latency, KPIs for critical IoT are the most stringent whereas those for massive IoT are the most relaxed. A-IoT is expected to have even more relaxed KPIs.
\section{A-IoT Use Cases and Requirements}
\label{sect:use_cases}
\subsection{Technology Overview}
\begin{figure*}
\centering
  	\includegraphics[width=5.5in]{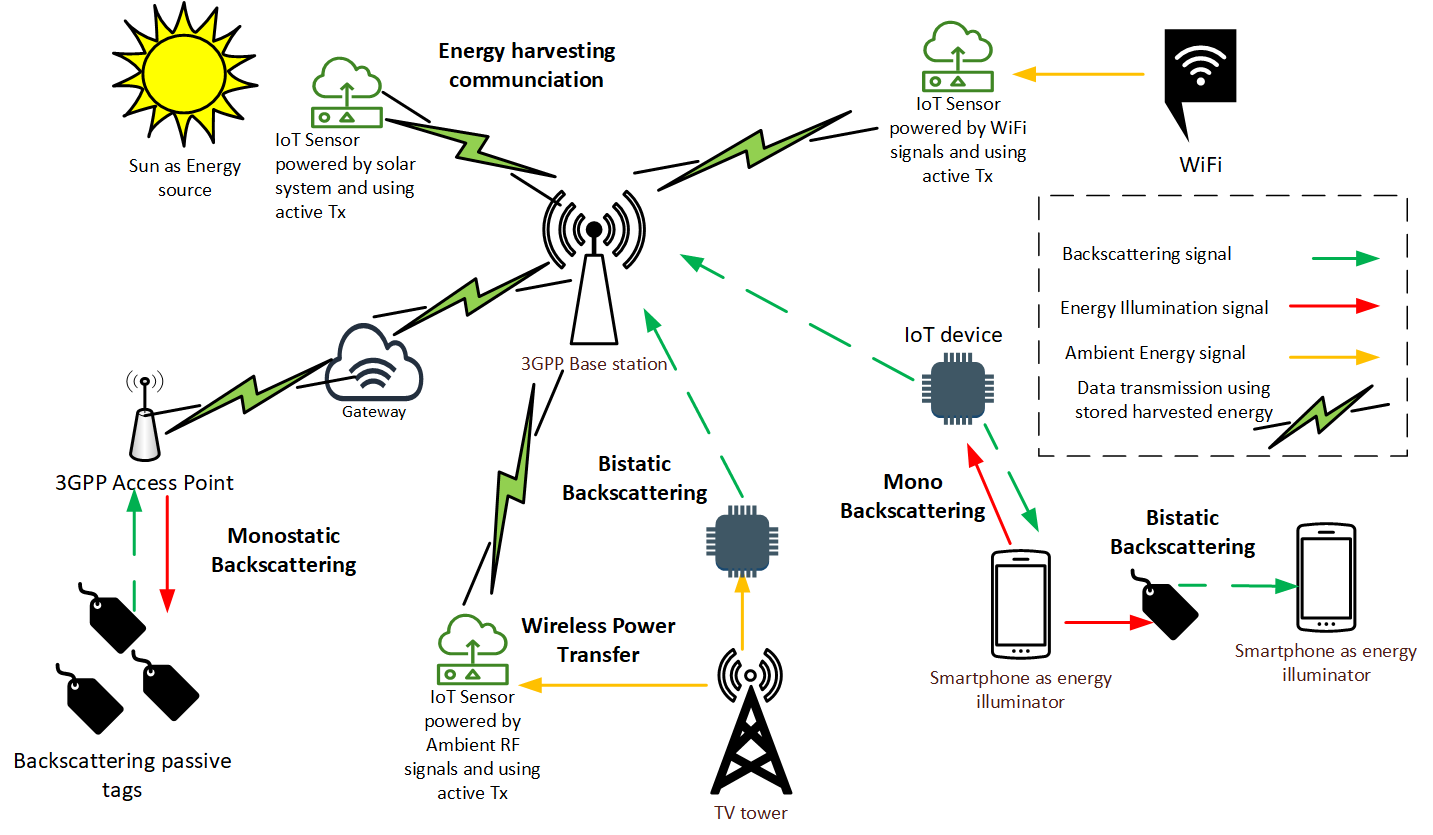}
   \caption{Ambient IoT landscape with both energy harvesting and backscattering devices. Ambient energy sources like sun, TV signals and Wifi signals can be used for energy harvesting and excitation as well.}
	\label{fig:intro}
\end{figure*}

A-IoT technology includes both energy harvesting devices with active transmission as well as passive backscattering devices. Backscattering devices do not have an active transmission component and modulate information on the received signal from the exciter where the exciter (often used interchangeably with the terms illuminator or activator) can be any node generating RF signals. On the other hand, energy harvesting with active transmission and small storage allows higher range and better quality of service (QoS) as compared to backscattering devices without any storage. Different use cases can be envisioned for both technologies, where low-cost backscattering technology is more suitable to use cases such as asset tracking and monitoring, livestock, etc., while high-end use cases requiring better QoS can be supported by energy harvesting technology with active transmission. Both backscattering and active energy-harvesting devices can be battery-less or carry a small battery.

Fig.~\ref{fig:intro} shows an overview of broad coverage of A-IoT technology. Backscattering IoT devices can be used both in monostatic and bi-/multi-static configurations, where the term monostatic is used when both the energy exciter and the reader functionalities are performed by the same device, while bistatic deployment specifies the scenario where the energy exciter and the reader are physically two different devices. Fig.~\ref{fig:intro} illustrates that 3GPP base stations (BSs, i.e., gNB or dedicated micro cells) and smartphones can be used both as exciters, and readers, or RF energy source, while an ambient energy source (e.g., TV, WiFi signals) can also be used for excitation for backscattering as well as energy source for energy harvesting for active transmissions. Besides RF signals, energy harvesting devices can make use of other ambient energy sources like sun light, pressure, thermal etc, to harvest and store energy and use it to make data transmission. Though 3GPP focus is more on RF energy harvesting devices in A-IoT studies, devices powered by other energy sources, e.g., solar which has several orders of magnitude high power density as compared to RF source,  are not precluded in this study.”

\subsection{3GPP Focus Areas and Use Cases}
\label{subsect:3gpp}
3GPP has recently  studied A-IoT in both Service and System Aspects working group 1 (SA1) \cite{AIoT_SA} and Radio Access Networks (RAN) plenary \cite{AIoT_RANTR}. The 3GPP work was initiated by the Technical Report in TR 22.840 \cite{AIoT_SA} by SA1 to capture use cases, traffic scenarios, device constraints of ambient power enabled IoT; and identify new potential service requirements as well as new KPIs.

Considering the limited size and complexity affordable by practical applications for battery-less devices with no energy storage capability or devices with limited energy storage that do not need to be replaced or recharged manually, the output power of the energy harvester typically ranges from 1 $\mu$W to a few hundreds of $\mu$W. Existing cellular devices may not work well with energy harvesting due to their peak power consumption of higher than 10 mW.

The RAN plenary later agreed to study A-IoT radio use cases and requirements in \cite{AIoT_RANTR}. This study has completed in 3GPP and a summary of main agreements is captured in this section.

3GPP RAN has agreed to study four deployment topologies for A-IoT devices:
\begin{enumerate}
  \item BS $\leftrightarrow$ A-IoT device
  \item BS $\leftrightarrow$ Intermediate node $\leftrightarrow$ A-IoT device
  \item BS $\leftrightarrow$ Assisting node UE $\leftrightarrow$ A-IoT device $\leftrightarrow$ BS
  \item UE $\leftrightarrow$ A-IoT device
\end{enumerate}
In Topology 1, the A-IoT device directly and bidirectionally communicates with a base station. In Topology 2, the A-IoT device communicates bidirectionally with an intermediate node between the device and the base station. The intermediate node can be a relay, Integrated Access and Backhaul (IAB) node, user equipment (UE), repeater, etc., which can support A-IoT. In Topology 3, the A-IoT device transmits data/signaling to a base station and receives data/signaling from the assisting node; or the A-IoT device receives data/signaling from a base station and transmits data/signaling to the assisting node. In this topology, the assisting node can be a relay, IAB, UE, repeater, etc. which can support A-IoT. In Topology 4, the A-IoT device communicates bidirectionally with a UE. 3GPP SA1 has defined a large set of use cases for ambient power-enabled IoT and 3GPP RAN is defining representative deployment scenarios for studies, each covering more use cases and topologies.

Using bi-/multi-static links can enable positioning and will remove the challenges associated with full duplexing self-interference for a monostatic A-IoT illuminator and reader device.

The 3GPP RAN study assumes three A-IoT device types:
\begin{itemize}
  \item \textbf{Device A:} (Passive) Pure battery-less devices with no energy storage capability at all, no independent signal generation/amplification (i.e., capable of only backscattering), and completely dependent on the availability of an external source of energy.
  \item \textbf{Device B:} (Semi-Passive) Devices with limited energy storage capability that do not need to be replaced or recharged manually, no independent signal generation but backscattering potentially with reflection gain.
  \item \textbf{Device C} (Active) Actively transmitting device with limited energy storage capabilities based on ambient energy sources.
\end{itemize}
\begin{figure}
\centering
  	\includegraphics[width=3.5in]{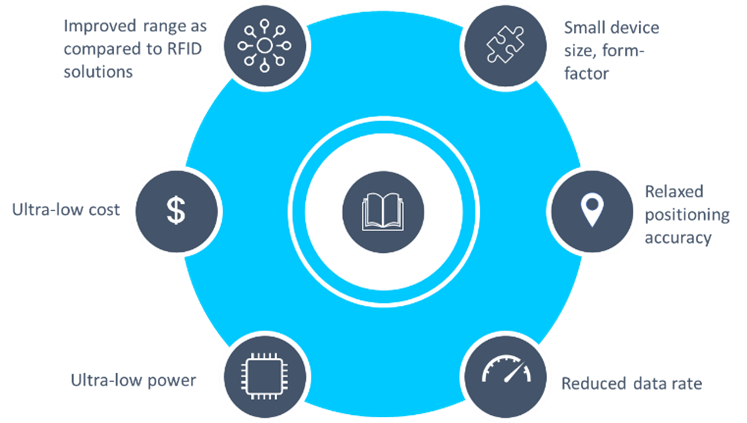}
   \caption{Design targets for A-IoT devices as discussed in 3GPP.}
	\label{fig:design}
\end{figure}

\subsection{Design Targets for Cellular AIoT Devices}
\begin{figure*}
\centering
  	\includegraphics[width=5.5in]{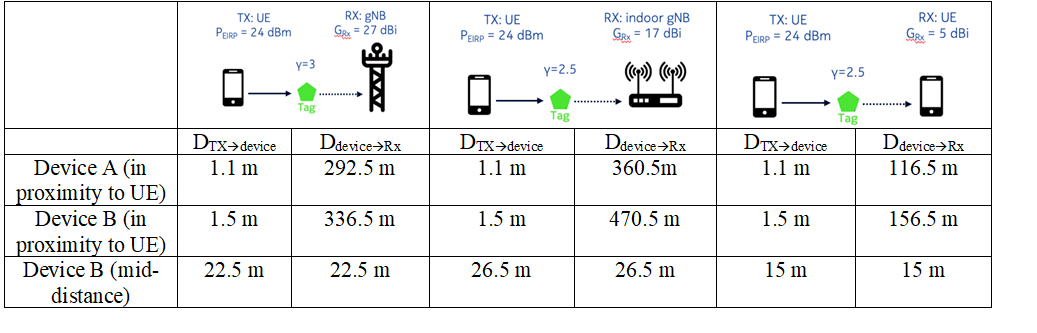}
   \caption{UE illumination scenarios with different receive options for device type A and B.}
	\label{fig:table}
\end{figure*}
The main design target pillars in focus for 3GPP A-IoT are depicted in Fig.~ \ref{fig:design}. The target is low data rate, ultra-low cost, ultra-low-power devices at small form factor but with improved range compared to Radio Frequency Identification (RFID) and with positioning enabled at relaxed accuracy.

The 3GPP RAN study assumes some high-level design targets for the A-IoT devices as outlined below.\\
\textbf{Device complexity:}
\begin{itemize}
  \item For Device A, the complexity target is to be comparable to ultrahigh frequency (UHF) RFID ISO18000-6C (EPC C1G2).

  \item For Device B, the target is such that: Device A complexity $<$ Device B complexity $<$ Device C complexity.
  \item For Device C, the complexity target is to be orders-of-magnitude lower than NB-IoT.
\end{itemize}
\textbf{Device power consumption:}
\begin{itemize}
  \item For Device A, the power consumption target during transmitting/receiving is $\leq$ 1 $\mu$W.
  \item For Device B, the target during transmitting/receiving is such that:
\begin{itemize}
\item Device A power consumption $<<$ Device B power consumption $<$ Device C power consumption; or
\item Device A power consumption $\leq$ Device B power consumption $<$ Device C power consumption.
\end{itemize}
  \item For Device C, the device power consumption is $\leq$ 1 mW.
\end{itemize}
\textbf{Device data rates:}\\
Maximum supported data rate not less than 5 kbps and minimum supported data rate not less than 0.1 kbps.\\
\textbf{Positioning accuracy (SA1 study based):}
\begin{itemize}
  \item Indoor: 3m at 90$\%$
  \item Outdoor: several 10m at $90\%$
\end{itemize}

Based on these simple design targets the ambition is to standardize radio interface and protocols to enable deployment of such Ambient IoT devices. 3GPP is expected to study  A-IoT devices in Rel. 19 with the aim to standardize in later releases. It is largely expected that AIoT technology will be one of the key bridge topics in 3GPP for 6G.

\section{Feasibility Studies}
\label{sec:sim}
As 3GPP studies are heavily focused on backscattering A-IoT devices, we discuss some insights on range coverage aspects for Devices A and B in this section. Through numerical evaluation on link budget for both forward and reverse links, we show which topologies are more practical, when taking into account the device type, transmit power level, and fixed vs mobile deployment. The received power $P_{\rm rx,tag}$ at the A-IoT device from a given transmitter is computed by product of two factors A and B such that $P_{\rm rx,tag}=AB$,
where $A=P_TG_T/d_1^\gamma$ and $B=G_{\rm tag}(\lambda/4\pi)^2$ \cite{AIoT_linkbudget}. $P_{\rm rx,tag}$ is the received power at a tag, $P_T$ is output power from a transmitter, $G_T$ is transmit antenna gain, $G_{\rm tag}$ is the tag antenna gain, $\lambda$ is the carrier wavelength, $d_1$ is the transmitter to tag distance and $\gamma$ is the path loss exponent ($\gamma=2$ for free space, $\gamma=3$ for fading environment). Similarly, at the reader, the received backscatter signal power $P_{\rm rx,r}$ is computed by $P_{\rm rx,r}=AB^2G_RM/d_2^\gamma$,
where $G_R$ is the receive antenna gain, $M$ is the backscatter tag modulation factor, and $d_2$ is the tag to receiver distance. Note that the above formula is generalized and applies to all backscatter deployment scenarios. For a monostatic case, it is imperative that $d_1=d_2$ and $G_T$ may be equal to $G_R$. For a bistatic case where the transmitter and receiver use equal antenna gain (e.g., two access points) the formula can be simplified by setting $G_T=G_R$. For a bistatic case where different equipment is used for transmitting and receiving (e.g. a UE and a BS), all parameters can be set individually.

Considering a tag performing binary modulation with a bit rate of 10 kbps and a target (uncoded) bit error rate (BER) = 1$\%$, the required backscatter signal to noise ratio (SNR) at the Rx is ${\rm SNR}_{\min} = 4.3$ dB. Assuming a 15 kHz signal bandwidth $W$, a receiver noise figure $NF = 6$ dB, a shadowing/fade margin of $F=10$ dB, the reader reference sensitivity has a value of approximately $S = -112$ dBm (i.e., using $S=-174+ NF+ F+10\log_{10}W+\rm {SNR}_{\min}$).

For battery-less devices that harvest RF energy (Device A), two conditions must be met for successful communication:
\begin{enumerate}
  \item Received excitation power at the tag $P_{\rm {tag}}$ must exceed a certain power-up threshold $P_{\rm thr}$ in order for the tag to "wake up". In the following calculations, a threshold of $P_{\rm {thr}} = -19$ $\rm {dBm}$ is assumed (common value for passive RFID devices). This typically governs the "max operating range".
  \item Received backscattered power at the receiver must exceed the reader sensitivity requirement $S$ ($S$ = -112 dBm in the following calculations).
\end{enumerate}

Three different bistatic use cases with UE excitation are depicted in Fig.~\ref{fig:table} with some example communication ranges. The path loss exponent $\gamma$ is assumed to be 3 for outdoor cases and 2.5 for indoor cases. For passive A-IoT devices (Device A), the modulation factor is $M$=0.25 and for semi-passive A-IoT devices (Device B) $M$=1. Note that all devices are assumed to have an omni-directional antenna ($G_{\rm {tag}}$ = 2 dBi), which represents a baseline scenario. In the first case from the left, excitation of devices is performed in an area by a UE and reception is performed by a fixed BS. The middle case involves indoor small-cell deployment with excitation of devices is performed by a UE and reception by a small-cell access point. The case to the right involves indoor small-cell deployment with device excitation and reading by two different UEs.

\begin{figure*}
\centering
  	\includegraphics[width=5.5in]{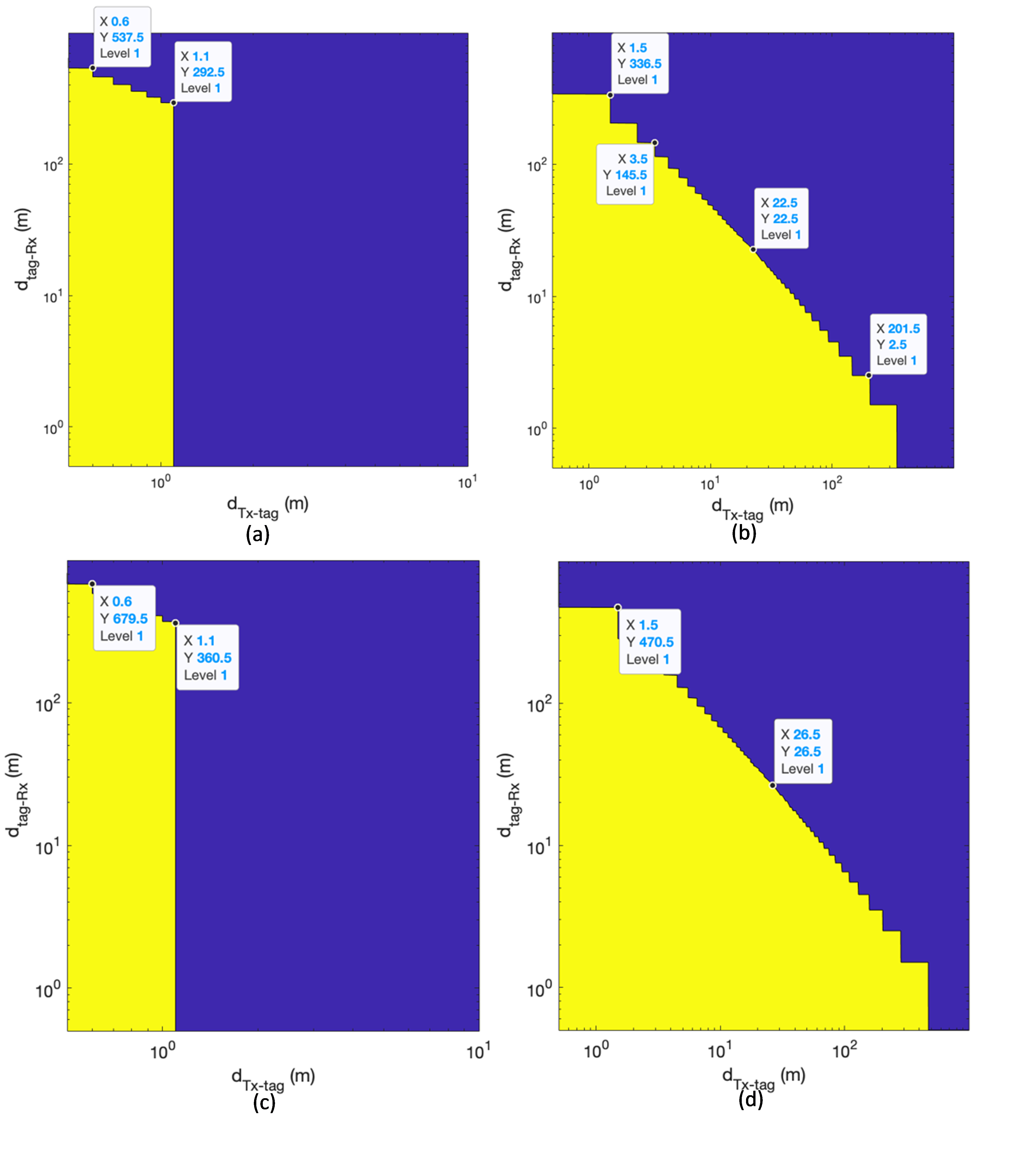}
   \caption{Operating regions (Yellow indicates sufficient received SNR) (a) for UE-device-gNB scenario device A operation (left) is governed by power-up threshold (b) for UE-device-gNB scenario device B operation (right) is receiver-sensitivity limited (c) for UE-device-indoor gNB scenario device A operation (left) is governed by power-up threshold (d) for UE-device-indoor gNB scenario device B operation (right) is receiver-sensitivity limited.}
	\label{fig:results}
\end{figure*}

It is interesting to note that for bistatic cases, the two links' (exciter (Tx)-to-device and device-to-receiver (Rx)) maximum range depends on each other. For a short Tx-device distance, a significantly higher device-Rx distance can be achieved. This has also been shown experimentally in \cite{AIoT_Kimionis}. The opposite is also true for semi-passive devices (Device B), but not for passive A-IoT devices that need to be energized by the exciter. Fig.~\ref{fig:results}(a,b) shows this contrast for the UE excitation-gNB reception case, where a yellow shaded region corresponds to successful communication operation. The Figure \ref{fig:results}(a) corresponds to an A-IoT device (Device A), for which the Tx-device distance is very limited, since the device power-up threshold must be exceeded. Despite that fact, very long device-Rx ranges can be achieved (several hundreds of meters) for a sufficiently sensitive receiver. Fig.~\ref{fig:results}(b) corresponds to an energy-assisted backscattering device (Device B). Since there is no RF power-up constraint, the total operating region is larger compared to Device A. Keeping a short Tx-device range allows for exponentially larger device-Rx range and vice versa. Similar trends are observed for the case of UE exciter and indoor gNB reception (Fig.~ \ref{fig:results}(c) and Fig.~\ref{fig:results}(d)).

Fig. \ref{fig:table} provides example numbers for the feasibility of each deployment (RF harvesting devices (Device A) and energy-assisted devices (Device B)). In general, Device B can afford longer Tx-device ranges than Device A, due to the relaxed power-up requirements, and therefore can be used for larger area coverage.
For UE excitation cases, where the UE can be brought to the proximity of the devices, both Device A and Device B can be used, and the uplink range from the device to a base station can reach 300 meters, which can accommodate similar inter-site distances (ISD).

Considering the above link budgets, the topology of illuminator and receiver to be deployed will depend on application and the environment. However, it is clear that for large scale outdoor or enterprise deployments, UE illumination will be favorable, since the power source can be brought close to the AIoT device. Then uplink ranges of several hundred meters can be achieved, and the AIoT device backscattering can be received by fixed infrastructure gNBs.

\section{A-IoT: TECHNOLOGY CHALLENGES AND SOLUTIONS}
\label{sect:future}
\subsection{A-IoT Device Access Techniques}
Energy harvesting IoT devices have the following key difference as compared to NR UE and associated challenges must be resolved before deploying A-IoT technology.
\begin{enumerate}
  \item {{\textbf{New Terminal States: }}A major consideration for the access techniques in energy harvesting communication is the uncertainty in the availability of energy. Unlike conventional communication scenarios, availability of ambient energy is not stable, and it varies spatially as well. When designing access techniques, this uncertainty has to be considered \cite{Majid_EH_patent}. For energy harvesting devices with active transmission (Device C), duty cycle design, scheduling decisions and wakeup occasions need to be designed based on availability of intermittent energy. As devices are not always connected with the network, NR functionality based on radio resource control (RRC) states (active/idle) may not be valid and we may need new device state definitions.}
  \item {\textbf{Device Registration: }} NR UEs are registered with the network and a context is maintained in the network. As A-IoT devices are low-cost devices with small form factor, it will be difficult to register them with the network through subscriber identity module (SIM). However, it is important to establish a simplified form of A-IoT device identification in a 3GPP network and 3GPP SA2 studies will look into this aspect. It is not yet clear how A-IoT device identification will be managed, i.e., through Subscription Permanent Identifier (SUPI) by the network or through application-defined IDs managed by a 3GPP network.
  \item{\textbf{Mobility Tracking: }}Typical use cases targeted by AIoT technology, e.g., inventory tracking, may not require continuous mobility management as in NR. Mobility management requires maintaining context of UE in the network and preparing handover when UEs move between cells. As some of the target use cases do not require this functionality, this overhead can be avoided by finding solutions for 'on demand' mobility management.
  \item {{\textbf{Link Budget Analysis: }}Backscattering devices are low-cost devices and typically first harvest some energy to activate their circuit and then backscatter information. Thus, 3GPP system design not only involves backscattering link budget analysis but illuminator-to-backscattering device analysis as well.}

  \item {{\textbf{Distributed Energy Sources: }}As shown in Section IV, due to a large link budget requirement, gNB is not the best exciter for backscattering devices. In 3GPP networks, all the NR UE access mechanisms are handled by gNB. For backscattering devices, we need to rely on distributed low-complexity 3GPP network illuminators, readers and/or smartphones, which require more coordination between network devices to service A-IoT devices and demand for new network protocol design.}

  \item {{\textbf{Simplified Protocols: }}Due to small form factor and low-cost requirement, A-IoT devices cannot support full stack access protocol. Thus, simplified protocol design is a key requirement for such devices, particularly backscattering devices which depend on strong activation signals. Access and security mechanisms need a new design approach that provides required authentication and security to backscattering devices at lower complexity and energy cost. Access and security protocol complexity may depend on each device type discussed in 3GPP.}
\end{enumerate}
\subsection{A-IoT Device Positioning Techniques}
\begin{figure}
\centering
  	\includegraphics[width=3.5in]{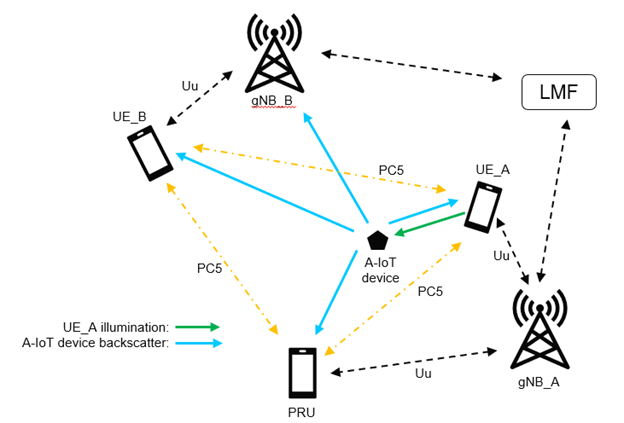}
   \caption{AIoT positioning incorporating various network entities. UE to UE links (PC5 interface) and UE to gNB links (Uu interface) have been marked.}
	\label{fig:positioning}
\end{figure}
For several 3GPP A-IoT deployment scenarios and use cases, support for positioning is a key feature. Fig.~\ref{fig:positioning} shows an example of an A-IoT deployment based on multi-static backscattering with positioning capabilities. The positioning of A-IoT devices can be based on the existing 3GPP positioning architecture, using a 3GPP node to trigger the event and one or more 3GPP nodes to listen to a given reference signal, e.g., another gNB or Positioning Reference Unit (PRU). The positioning session can be controlled and monitored by the location management function (LMF). A positioning session can be enabled by a UE in the network or from the network side. Sidelink can be another means to configure UEs in proximity of the A-IoT device as activation exciter and/or as positioning anchors.

A-IoT positioning encompasses two complementary use-cases:  1) locate the A-IoT device and 2) use the A-IoT device to locate other devices.

In the first use case, the A-IoT device is often used for tracking objects in production, shipping, etc., and therefore, it is necessary for the NR network to locate the A-IoT device. However, since the A-IoT activation and reading ranges and device processing capabilities are severely limited, the typical time- and angle-based NR positioning techniques cannot be straightforwardly applied. Specifically, A-IoT systems do not lend themselves well to downlink positioning techniques, since these requires the A-IoT device to be synchronized to the network and be able to detect positioning signals and extract relevant measurements (differential timing measurements, angle of departures, carrier phases, etc.), requirements which are impossible to fulfill by most A-IoT devices. Round-trip time (RTT) methods pose an even more challenging task, since the A-IoT device must, in addition to processing downlink, also reply with a paired uplink transmission. Given the above, the only feasible alternative remains uplink positioning, where the Transmission/Reception Points (TRPs) are tasked with detecting the device and measuring the required positioning metrics, conditioned on the availability and proximity of radios to charge and activate the A-IoT device.

In the second use-case, the A-IoT device may itself be used to locate other devices (other A-IoT devices or NR UEs). Specifically, NR localization of a UE requires signaling from multiple TRPs with known location, to multilaterate the UE location. When an insufficient number of TRPs are available and/or in poor line-of-sight (e.g., very often in indoor scenarios), cheap A-IoT devices may take the role of the TRP, and become the so-called A-IoT TRP. However, the location of an A-IoT TRP:
\begin{enumerate}
  \item {is typically unknown to the NR network right after deployment. That is because such devices are manually installed at physical locations associated with areas which have positioning blind spots.}
  \item {may change over time. The devices may be manually moved around to cover other/new positioning blind spots e.g., where such blind spots are created when an existing TRP gets obstructed due to temporary blockage. For example, a TRP can be blocked in an indoor factory when large packages are moved around the factory or new equipment is installed.}
  \item {cannot be computed using NR positioning techniques due to the limited range of such devices.
While it has high cost-reduction potential, it is not straightforward how to efficiently utilize an A-IoT device as a TRP, since the current NR positioning protocols are not equipped to compute and track the position of the A-IoT device.}
\end{enumerate}

\subsection{A-IoT Authentication and Security}
Similar to RFID, security methods for A-IoT systems must guarantee at least (a) authentication, i.e., confirming device identity and (b) confidentiality, i.e., ensure that the A-IoT device is not being eavesdropped on. NR security mechanisms guarantee both requirements, data integrity and availability via higher layer protocols and complex cryptography. The complexity of NR security is however both a blessing (for existing services) and a curse for A-IoT systems. A-IoT systems consist of devices that are extremely limited in terms of computational power, that can rarely support basic cryptography. Therefore, there is a need to develop new security protocols and methods tailored to A-IoT design limitations. The above-mentioned protocols may instead resort to physical layer security mechanisms. Physical layer security is built on exploiting the uniqueness of the propagation channel responses between the A-IoT device and other legitimate radios comprising the A-IoT system and determining a transmission scheme tailored to those specific conditions only.
\subsection{Spectrum}
The spectrum to be targeted for A-IoT should be common across as many regions as possible and with the same or similar spectrum access regulations, both to ensure the ease of mobility of AIoT devices across different regions as well as to simplify the standardization of the A-IoT ecosystem. For example, RFID technology can be deployed across different bands, where the actual band depends on the use case and the requirements \cite{AIoT_spectrum}. However, in order to avoid coexistence issues with legacy RFID technologies and at the same time take advantage of the existing cellular deployments, the A-IoT system should target licensed frequency bands where cellular technologies already operate as well as the unlicensed bands at 5 GHz and 6 GHz.
In particular, the unlicensed bands will have both infrastructure-based and sidelink-based access technologies allowing for the coordination of the A-IoT communications and aid RF-based charging of the A-IoT devices.


\section{Conclusion}
\label{sect:conclusions}
This article provides a comprehensive overview of Ambient IoT technology in 5G. Ambient IoT is new technology on top of existing cellular IoT technology and targets market segment below the existing technologies. Battery replenishment and cost is major hurdle in deployment of IoT technology and A-IoT can help fill this gap by targeting reduced complexity and cost. Before this technology becomes a reality, several challenges need to be overcome including development of suitable access, transmission, positioning, and radio resource management techniques, identifying spectrum and topologies for various scenarios, and incorporating low complexity security protocols. This article summarizes 3GPP discussions on A-IoT studies and envision an ambient IoT technology that will make future networks more sustainable and ready for massive deployment of IoT devices.

\bibliographystyle{IEEEtran}
\bibliography{bibliography}
%
%
%


\end{document}